\documentclass{mem}
\usepackage{natbib}\usepackage{txfonts}\usepackage{balance}
\usepackage{graphicx}
\usepackage[a4paper]{hyperref}
\idline{75}{282}
\begin{document}
\def\teff{$T\rm_{eff }$}
\def\kms{$\mathrm {km s}^{-1}$}
\def\cgs{ ${\rm erg~cm}^{-2}~{\rm s}^{-1}$ }    
\def\es{ ${\rm erg~\rm s}^{-1}$ }    
\def\gtrsim{\mathrel{\hbox{\rlap{\hbox{\lower4pt\hbox{$\sim$}}}\hbox{$>$}}}} 

\title{
Heavily obscured AGN with BeppoSAX, INTEGRAL, SWIFT, XMM and Chandra: 
prospects for 
Simbol-X
}

   \subtitle{}

\author{ 
R. \,Della Ceca \inst{1}, 
P. \,Severgnini \inst{1}, 
A. \,Caccianiga \inst{1},
A. \,Comastri \inst{2},
R. \,Gilli \inst{2},
F. \,Fiore \inst{3},
E. \,Piconcelli \inst{3},
P. \,Malaguti \inst{4},
and 
C. \,Vignali \inst{5,2},
}

  \offprints{R. Della Ceca, E-mail: roberto.dellaceca@brera.inaf.it}

\institute{
Istituto Nazionale di Astrofisica --
Osservatorio Astronomico di Brera, Milano, Italy
\and
Istituto Nazionale di Astrofisica --
Osservatorio Astronomico di Bologna, Italy
\and 
Istituto Nazionale di Astrofisica --
Osservatorio Astronomico di Roma, Italy
\and
Istituto Nazionale di Astrofisica --
IASF, Bologna, Italy
\and
Dipartimento di Astronomia, Universita´ di Bologna, Italy
}

\authorrunning{Della Ceca}

\titlerunning{Heavily Obscured AGN}

\abstract{
According to the latest versions of synthesis modeling of the Cosmic X-ray
Background,  Compton Thick AGN  should represent $\sim 50\%$ of the total
absorbed AGN population. However, despite their importance in the cosmological
context, only  a few dozens of Compton Thick AGN have been found and studied so
far. We will briefly review this topic and discuss the improvement in
this field offered by the Simbol-X mission with its leap in sensitivity 
(E$>10$ keV) of more than a factor 500 with respect to previous X-ray missions.
}

\maketitle{}

\section{Introduction}

Active Galactic Nuclei (AGN) emit over the entire electromagnetic spectrum and
are widely believed to be powered by accretion of matter onto a Supermassive
(millions to billions solar masses) Black  Hole (SMBH; \citealp{rees84}). It is
now well established that the largest fraction of the AGN population  is
obscured by a large amount of cold matter around the Active Nuclei that does not
permit a direct view to the central energy source.   The identification of
sources hosting such obscured accreting nuclei is a difficult task: in the
optical domain the active nuclei appear very dim and their luminosity is
comparable to that of their host galaxies (leaving very weak indications of the
presence of an AGN), while in the  X-ray band (up to 10 keV) their selection is
difficult as even hydrogen column densities  ($N_H$) of the order of
$10^{22}$-$10^{23}$ cm$^{-2}$ strongly reduce the flux from the nuclear source. 

However, despite their elusiveness, obscured sources are fundamental for our 
understanding of the SMBHs history as the large  majority of the energy density
generated by accretion of matter in the Universe seems to takes place in obscured
AGN (\citealp{fabian98}), as testified by the  integrated  energy density
contained in the cosmic X-ray background  (XRB;  \citealp{gilli07}, Comastri et
al., these proceedings). Even more important,  the recent discovery of quiescent
SMBH in the nuclei of non-active nearby galaxies with prominent bulges (\citealp{kormendy95}, 
\citealp{magorrian98}), along with the presence of scaling relations between the
central BH mass and galaxy  properties (e.g. buldge luminosity/mass  and
velocity dispersion, \citealp{ferrarese00})  strongly suggest that AGN are leading
actors in the  formation and evolution of galaxies and, in general, of cosmic
structures in the Universe (see \citealp{begelman04} and references therein).

For absorbed AGN having column densities up to few times $10^{23}$ cm$^{-2}$
XMM-Newton and {\it Chandra} have already produced (and are still producing) a
wealth of useful data up to $\sim$10 keV that can be used to evaluate their
statistical properties as a function of cosmic time  (e.g. \citealp{lafranca05},
\citealp{mueller07}, Della Ceca et al., in preparation). On the other hand, the
situation is almost completely unconstrained for the absorbed sources  having
$N_H$ in excess of $\sim 10^{24}$ cm$^{-2}$ (the so called Compton Thick AGN),
whem the matter is optically thick to Compton scattering. According to the
latest versions of synthesis modeling of the  XRB (\citealt{gilli07}), these
sources should represent $\sim 50\%$ of the total absorbed AGN population but
only a few dozens of them (mostly local) have been found and studied so far.

We will briefly review here what do we know about the Compton Thick AGN 
Universe and which is the improvement offered by the Simbol-X mission. A review 
of several aspects  of the Compton Thick AGN field can be found in
\cite{comastri04}.  Throughout this paper we consider
($H_o$,$\Omega_M$,$\Omega_{\Lambda}$)=(70,0.3,0.7).

\section{The Compton Thick AGN Universe}

The Compton Thick AGN may be separated into two classes.
If the absorbing column density is below a few times  $10^{24}$ cm$^{-2}$, a
significant fraction of the the high energy  (E$>$10 keV) photons can escape
after one or more scatterings, allowing the  nuclear source to be directly
visible above 10 keV.  In this case the source is called mildly Compton Thick 
(a good example is NGC 4945, \citealp{guainazzi00}, see also section 3 for a
typical X-ray  spectrum). On the contrary, if the absorbing column densities is
close or above  $10^{25}$ cm$^{-2}$, then the X-ray photons cannot escape even
in hard X-rays,  being trapped in the matter, downscattered to lower energies
and  eventually destroyed by photoelectric absorption; the X-ray spectrum is
hence  depressed over the entire X-ray energy range.  In this case the source is
called heavily Compton Thick; a well known example is NGC 1068 
(\citealp{matt97}).  The presence of Compton thick matter may be inferred
through indirect arguments, such as the presence of a strong Iron K$_\alpha$
line complex at 6.4 - 7 keV and the characteristic reflection spectrum (having
the emission peak above 10 keV). It is thus clear that hard X-ray data above 10
keV are  fundamental to unveil and study Compton Thick AGN and to have a real
and unbiased census of SMBH in the Universe.

\begin{figure}[]
\resizebox{\hsize}{!}{\includegraphics[clip=true]{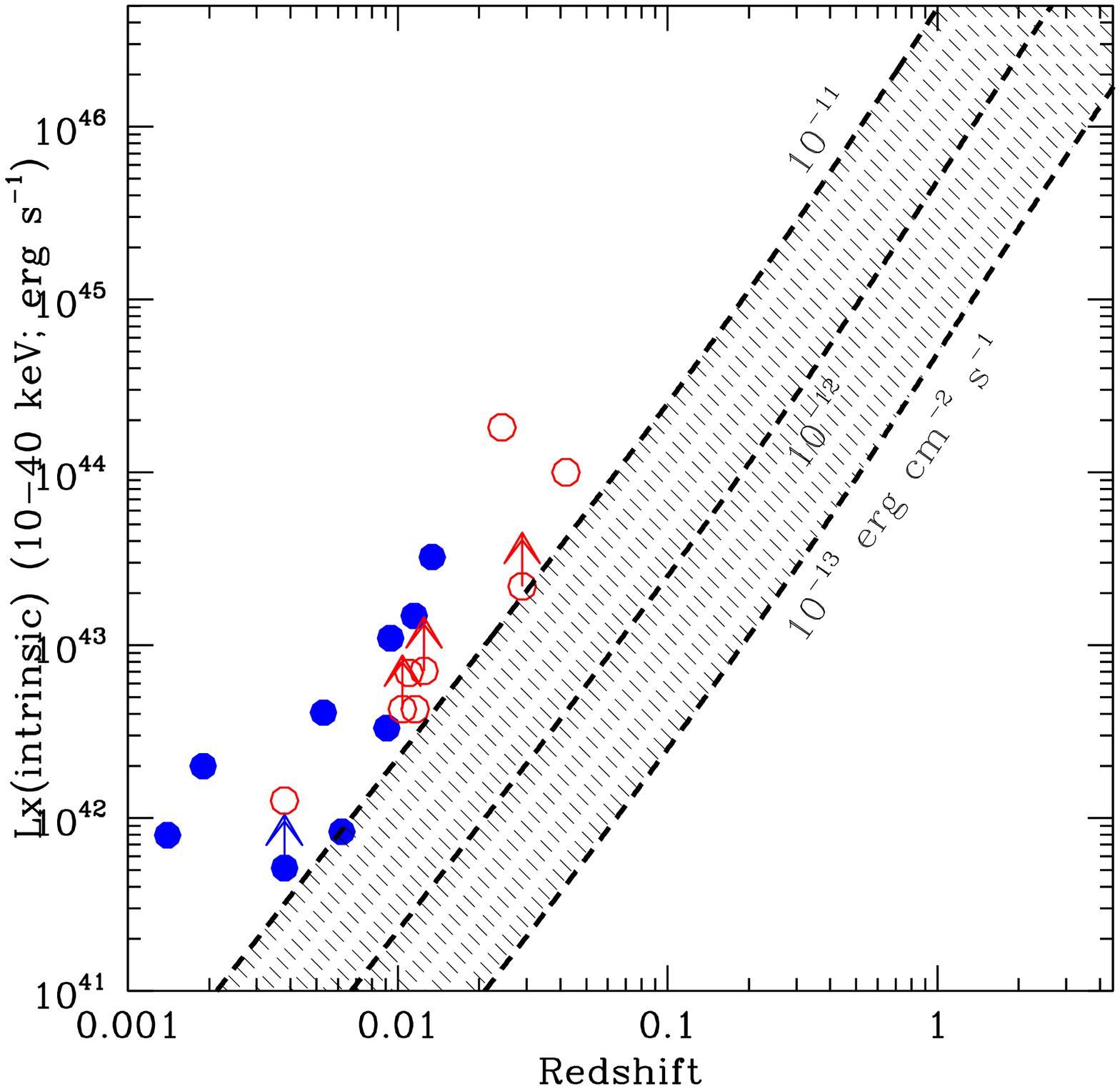}}
\caption{
\footnotesize
Position in the luminosity-redshift plane of the 18 {\it bona fide} Compton
thick AGN detected  so far above 10 keV to our knowledge. The arrows mark the
heavily Compton thick sources for which even the measured luminosity above 10
keV represents a lower limit to the intrinsic ones. The dashed lines represent
the flux limit (10--40 keV energy band) 
of  $10^{-11}$, $10^{-12}$ and $10^{-13}$
\cgs.  The filled circles mark the objects belonging to complete and well
defined survey programs, while the open circles mark the sources coming from
pointed observatios of objects selected at other energies. The shaded area
enclose the discovery space offered by the  Simbol-X mission regarding the
spectral investigations of Compton Thick AGN  (see section 3). For MKN 231 we
have used a (10--40 keV) luminosity of $10^{44}$ \es.
}
\label{fig1}
\end{figure}

Our present knowledge of the Compton Thick AGN universe is mainly based on few
{\it bona fide} Compton Thick sources.
In Table 1 we have reported all the 18 AGN so far identified as 
Compton Thick sources (to our knowledge) thanks to observations above 10 keV 
with {\it Beppo}SAX, INTEGRAL, SWIFT/BAT and SUZAKU
\footnote{
Table 1 and Table 2 are an updated version of the  list of Compton
Thick AGN discussed in \cite{comastri04}.};
their position in the luminosity redshift plane is shown in  Fig.\ref{fig1}.  

All the  objects are at z$<$ 0.05
and the large majority  are relatively low luminosity ($<10^{43}$\es) sources,
a clear consequence of the flux limit above 10 keV ($\sim 10^{-11}$\cgs) of the
X-ray missions carried out so far. 
Nine of the Compton Thick sources belong to
well defined and complete samples selected above 10 keV (e.g. \citealp{sazonov07},
\citealp{beckmann06}, \citealp{bassani06}, \citealp{markwardt05}), while the remaining 9
sources come from specific  pointed observations of objects selected at other
wavelengths. Finally only 4 heavily Compton Thick  AGN 
(NGC 1068, NGC 7674, NGC 3393 ed NGC 4939) have been detected so far above
10 keV.

\begin{table*}
\caption{{\it Bona fide} Compton Thick AGN detected above 10 keV$^a$}
\label{abun}
\begin{center}
\begin{tabular}{lccccccc}
\hline
\\
Name          & z      & Opt.Class. & $N_H$                & Log$L_{10-40 keV}^b$  & Log$f_{2-10 keV}$ & EW$_{Fe}$           & Ref. \\
              &        &          & ($10^{22}$ cm$^{-2}$)  &  \es                   & \cgs               & (eV)                &      \\
\hline
\\
Circinus      & 0.0014 &  Sy2       & $430_{-70}^{+40}$    & 41.90            &  -10.85       & $2310^{+120}_{-260}$ &  6,2,3,4,7,24  \\
M51           & 0.0015 &  LIN/Sy2   & $560_{-160}^{+400}$  & 40.76            &  -12.47       & $900^{+400}_{-300}$  &  19          \\
NGC 4945      & 0.0019 &  LINER     & $220_{-40}^{+30}$    & 42.30            &  -11.27       & $\sim$ 1300          &  12,3,2,4     \\
NGC 1068      & 0.0038 &  Sy2       & $>$1000              & $>$41.71         & -11.41       & $\sim$ 3100          &  1,2,3,4,5,    \\
NGC 3079      & 0.0038 &  LIN/Sy2   & $1000_{-530}^{+1000}$ & 42.10           &  -12.42       & $2400^{+2900}_{-1500}$ &  20        \\
NGC 7582$^c$   & 0.0053 &  Sy2       & $160_{-46}^{+94}$    & 42.61            &  -10.71       & $570^{+570}_{-120}$  &  22,10       \\
SWIFT0601.9-8636    & 0.0062 &  Gal       & $101_{-38}^{+54}$    & 41.92            &  -11.96       & $1060^{+160}_{-160}$ &  25          \\
ESO138-G01    & 0.0091 &  Sy2       & $\sim$ 150           & 42.52            &  -11.74       & $1700^{+400}_{-400}$ &  4,14        \\
NGC 5728      & 0.0094 &  Sy2       & $210_{-20}^{+20}$    & 43.04            &  -11.82       & $1000^{+300}_{-300}$ &  23, 10      \\
NGC 4939      & 0.0104 &  Sy2       & $>$1000              & $>$ 42.63         &  -11.85       & $480^{+420}_{-210}$  &  21          \\
NGC 3690      & 0.0110 &  HII       & $252_{-56}^{+139}$   & 42.84            &  -11.95       & $636^{+236}_{-270}$  &  16,17       \\  
NGC 3281      & 0.0115 &  Sy2       & $196_{-5}^{+20}$     & 43.17            &  -11.54       & $526^{+128}_{-144}$  &  18,2,10     \\
Tol 0109-383  & 0.0116 &  Sy2       & $350_{-140}^{+180}$  & 42.63            &  -11.80       & $\sim$ 1360          &  13,14        \\
NGC 3393      & 0.0125 &  Sy2       & $>$1000              & $>$ 42.85         &  -12.41       & $1400^{+700}_{-700}$ &  21,5       \\
MKN 3         & 0.0134 &  Sy2       & $127_{-22}^{+24}$    & 43.51            &  -11.19       & $997^{+300}_{-307}$  &  9,10,3,2,4,7 \\    
NGC 6240      & 0.0243 &  LINER     & $218_{-27}^{+40}$    & 44.26            &  -11.71       & $1580^{+380}_{-350}$ &  7,8          \\
NGC 7674      & 0.0289 &  Sy2       & $>$1000              & $>$43.34         &  -12.30       & $900^{+470}_{-299}$  &  11           \\
MKN 231       & 0.042  &  BAL QSO   & $\sim 200$           & 43.7-44.3        &  -12.15       & $\sim 250$           &  26          \\
%**IRAS09104     & 0.442  &  QSO2      & $>$500             & $>$45.79          &   -           & $\sim$ 1900          &  15           \\
\\
\hline
\end{tabular}
\end{center}
References: 
(1) \cite{matt97};
(2) \cite{sazonov07};
(3) \cite{beckmann06};
(4) \cite{bassani06};
(5) \cite{levenson06};
(6) \cite{matt99};
(7) \cite{bassani99};
(8) \cite{vignati99};
(9) \cite{cappi99};
(10) \cite{markwardt05};
(11) \cite{malaguti98};
(12) \cite{guainazzi00};
(13) \cite{iwasawa01};
(14) \cite{collinge00};
(15) \cite{franceschini00};
(16) \cite{dellaceca02};
(17) \cite{ballo04}; 
(18) \cite{vignali02};
(19) \cite{fukazawa01};
(20) \cite{iyomoto01};
(21) \cite{maiolino98};
(22) \cite{turner00};
(23) \cite{comastri07};
(24) \cite{iwasawa97};
(25) \cite{ueda07};
(26) \cite{braito04}.

Table notes - 

$\ ^a$ We have not considered here the Compton Thick AGN candidate IRAS
09104+4109 (a type 2  QSO at z=0.442). The Compton Thick nature of this object
had been previously suggested (\citealp{franceschini00}) by a  marginal ($\sim 2.5
\sigma$) detection in the 15-50 keV energy range and  recently re-discussed (and
not confirmed)  by  \cite{piconcelli07} using XMM-Newton data; 

$\ ^b$ The 10-40 keV luminosities have ben computed assuming a 
power-law spectral model with photon index equal to 1.9 and, for the  mildly
Compton Thick AGN have ben corrected for absorption. For the sources  having 
$N_H \gtrsim 10^{25}$ cm$^{-2}$ this correction is not possible  and the computed
luminosities are a lower limit to the intrinsic ones;

$^c$ This source has been recently investigated by \cite{piconcelli07b} using
XMM-Newton data. The XMM-Newton spectrum of the nuclear source is very  complex
(a combination of a midly absorbed, $N_H \sim 10^{24}$ cm$^{-2}$, power-law and
a pure reflected component both obscured by a column density of $\sim 4\times
10^{22}$ cm$^{-2}$) also showing dramatic spectral changes. 

\end{table*}

At the few {\it bona fide} Compton Thick sources reported in Table 1 we may  
add the Compton Thick AGN {\it candidates}, for which we have observations only
below 10 keV. From these observations, the heavily absorbed active nature of
these sources is suggested by the presence of a strong (EW $\gtrsim 500$
eV) Fe emission line, a characteristic property of the Compton Thick AGN
population.  The Compton Thick AGN candidates identified so far in this way are
reported in Table 2.   We note that for these sources (almost all at very low
redshift) we lack the coverage above  $\sim$10 keV, thus we are unable to
confirm the presence of an absorption cut off in the hard X-ray domain; only a
conservative lower limit on the intrinsic $N_H$ and nuclear luminosity can be
placed.

About 30-50 Compton Thick AGN candidate have been also selected in the Chandra
deep fields on the basis of (poor quality)  X-ray and optical data (e.g.
\citealp{tozzi06}, \citealp{alexander03}) and a few objects  have been selected
using the Sloan Digital Sky Survey (\citealt{vignali06}); a few other candidates
have been recently selected using X-ray data combined with Spitzer data (e.g.
\citealp{polletta06}, \citealp{daddi07}, \citealp{fiore07}).   We stress again
that also for the Compton Thick candidates from the Chandra Deep fields and/or
from Spitzer the intrinsic absorption column density and the intrinsic
luminosities are unconstrained.

\begin{table*}
\caption{Compton Thick AGN candidates detected below 10 keV}
\label{abun}
\begin{center}
\begin{tabular}{lccccccc}
\hline
\\
Name             & z      & Opt.Class. & $N_H$                  & Log$f_{2-10 keV}$   & Log$L_{2-10 keV}^{a}$  & EW$_{Fe}$              & Ref. \\
                 &        &            & ($10^{22}$ cm$^{-2}$)  &  \cgs                &  \es                & (eV)                   &      \\
\hline
\\
NGC 1386         & 0.0029 &    Sy2     & $>$100            & -12.62         &  39.65           & $2300^{+1500}_{-1500}$ & 5,2    \\
NGC 5643         & 0.0039 &    Sy2     & $>$1000           & -11.95         &  40.57           & $1900^{+1400}_{-700}$  & 5      \\
NGC 1365$\ ^b$    & 0.0055 &    Sy2      & $>$100            & -12.03         &  40.79           & $2100^{+2100}_{-300}$  & 17      \\
ESO 428\_G14      & 0.0057 &    Sy2     & $>$150            & -12.42         &  40.44           & $1600^{+500}_{-500}$   & 2     \\
NGC 2273         & 0.0062 &    Sy2     & $>$1000           & -12.00         &  40.93           & $2490^{+800}_{-680}$   & 5      \\
NGC 5347         & 0.0078 &    Sy2     & $>$100            & -12.66         &  40.47           & $1300^{+500}_{-500}$   & 2,13   \\  
IC 2560          & 0.0097 &    Sy2     & $>$100            & -12.44         &  40.88           & $3600^{+1500}_{-1500}$ & 7     \\
NGC 4968         & 0.0099 &    Sy2     & $>$160            & -12.82         &  40.51           & $3000^{+1200}_{-1000}$ & 1     \\
UGC 2456         & 0.0120 &    Sy2     & $90^{+60}_{-30}$  & -12.44         &  41.06           & $1000^{+300}_{-400}$   & 1      \\
IC 3639          & 0.0110 &    Sy2     & $>$1000           & -12.46         &  40.97           & $3200^{+980}_{-1740}$  & 4     \\
MKN 1210         & 0.0135 &    Sy2     & $>$100            & -11.89         &  41.72           & $820^{+360}_{-430}$    & 6     \\
NGC 5135         & 0.0137 &    Sy2     & $>$100            & -12.80         &  40.82           & $1700^{600}_{-800}$    & 1     \\
NGC 591          & 0.0152 &    Sy2     & $>$160            & -12.70         &  41.01           & $2200^{+700}_{-600}$   & 1     \\
IC 4995          & 0.0161 &    Sy2     & $>$160            & -12.54         &  41.22           & $1700^{+700}_{-700}$   & 1     \\
UGC 1214         & 0.0172 &    Sy2     & $>$160            & -12.55         &  41.27           & $1300^{+500}_{-500}$   & 1      \\
NGC 6552         & 0.0262 &    SY2     & $>$100            & -12.22         &  41.97           & $\sim$ 900             & 6,14   \\
NGC 7212         & 0.0266 &    Sy2     & $>$160            & -12.16         &  42.04           & $900^{+200}_{-300}$    & 1,2    \\
MKN 266          & 0.0279 &    Sy2      & $>$100            & -12.25         &  41.99           & $\sim 575$             & 16      \\
UGC 5101         & 0.0394 &    LINER   & $140^{+10}_{-20}$ & -12.21         &  42.34           & $ 410^{+270}_{-240}$   & 8,9   \\
IRAS 19254-7245  & 0.062  &    HII     & $>$100            & -12.62         &  42.33           & $2000^{+600}_{-600}$   & 10    \\
S5 1946+708      & 0.1010 &    NELG    & $>$280            & -12.28         &  43.11           & $1800^{+1200}_{-1100}$ & 3      \\
XID 2608         & 0.125  &    Sy2      & $>$150            & -13.29         &  42.29           & $792^{+1151}_{-493}$   & 12    \\
IRAS00182-7112   & 0.327  &    LINER    & $>$100            & -12.82         &  43.70           & $910^{+460}_{-460}$    & 18      \\
IRAS F15307+3252 & 0.93   &    Sy2/QSO2    
                                      & $>$100            & -13.46         &  44.01           & $>$2000                & 11    \\
CDFS 202         & 3.7    &    Sy2/QSO2 & $>$1000           & -14.56         &  43.85           & $1185^{+1195}_{-922}$  & 15      \\
\\
\hline
\end{tabular}
\end{center}
References: 
(1) \cite{guainazzi05};
(2) \cite{levenson06};
(3) \cite{risaliti03};
(4) \cite{risaliti99};
(5) \cite{maiolino98};
(6) \cite{bassani99};
(7) \cite{iwasawa02};
(8) \cite{imanishi03};
(9) \cite{ptak03};
(10) \cite{braito03};
(11) \cite{iwasawa05};
(12) \cite{mainieri06};
(13) \cite{risaliti99b};
(14) \cite{reynolds94};
(15) \cite{norman02};
(16) \cite{risaliti00};
(17) \cite{iyomoto97};
(18) \cite{nandra07}.

Table notes --
$\ ^a$ Observed luminosity (i.e. not corrected for absorption);
$\ ^b$ According to \cite{risaliti05} this source shows rapid 
Compton-Thick/Compton-Thin transitions.

\end{table*}

\section{The Simbol-X view of Compton Thick AGN} 

The Simbol-X mission 
\footnote{The Simbol-X mission and its main scientific objectives are discussed in 
the F. Ferrando and F. Fiore contributions at this meeting.}
is expected to have a leap in sensitivity in the 10--40 keV band of more than 
a factor 500 with respect to previous X-ray missions.

To evaluate the impact of this mission  on the studies of Compton Thick AGN at
cosmological distances  we have run spectral simulations using the satellite
configuration and  response files as reported in the notes posted  in
http://www.iasfbo.inaf.it/simbolx/faqs.php (March 28th, 2007 version).

To this purpose we assumed an adequate template of Compton Thick AGN. The
adopted spectral model was a combination of the following   components: a) a
primary power-law component filtered by an absorbing column density 
$N_H>10^{24}$ cm$^{-2}$; b) a scattered component having a scattered efficiency
of $\sim 1\%$; c) a reflected component which accounts for few percent of the
total  unabsorbed flux in the 2-10 keV energy range and  d) an Iron line complex
(Fe lines at 6.4 and 6.97 keV) having an observed EW $\sim 1$ keV and a line
ratio fixed from  the atomic physics. All these spectral components are commonly
observed in Compton Thick AGN (e.g the  recent SUZAKU observation of NGC 5728
discussed in \citealp{comastri07}).

In Fig. \ref{fig2}, \ref{fig3}, and \ref{fig4} we report three simulations (100
ksec each) corresponding to a midly Compton Thick AGN  ($N_H\sim 3\times
10^{24}$ cm$^{-2}$) having  $f_{10-40 keV} \sim10^{-11}$, $\sim10^{-12}$ and
$\sim10^{-13}$ \cgs, while in  Fig. \ref{fig5} we show a 100  ksec Simbol-X
simulation  of a  heavily Compton Thick AGN  ($N_H > 10^{25}$ cm$^{-2}$) for
which we see only the reflected component and the Fe line complex.  These
simulations are briefly discussed below along with some considerations.

\subsection{$f_{10-40 keV} \sim 10^{-11}$ \cgs}

\begin{figure}[]
\resizebox{\hsize}{!}{\includegraphics[angle=-90,clip=true]{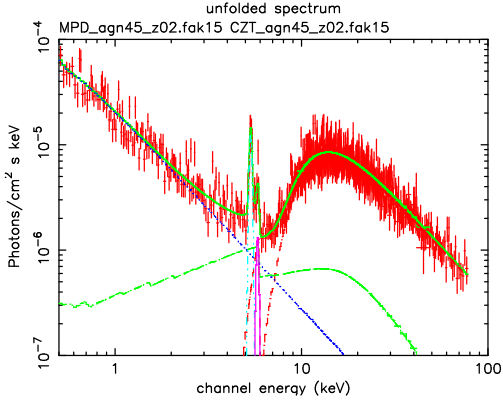}}
\caption{
\footnotesize
Simulated 100 ksec Simbol-X observation for a midly Compton Thick 
AGN ($N_H\sim 3\times
10^{24}$ cm$^{-2}$, spectral model as described in the text) with $f_{10-40
keV} \sim 10^{-11}$ \cgs. In particular this simulation  corresponds to a source
with a $L_{10-40 keV} \sim 10^{45}$ \es at z$\sim$ 0.2.  
}
\label{fig2}
\end{figure}

The simulation reported in Fig.\ref{fig2} corresponds to a Compton Thick AGN
with a $L_{10-40 keV} \sim 10^{45}$ \es at z$\sim$0.2. With a Simbol-X  exposure
of 100 ksec we should accumulate $\sim$ 11000 net counts  from the Macro Pixel
Detector (MPD) and $\sim$ 8000 net counts from the Cd(Zn)Te detector (CZT).  The
analysis of these data shows that the  $N_H$  will be determined  with an
accuracy (90\% confidence error) of $\sim 5\%$, while  the intrinsic luminosity
will be determined with an accuracy of $\sim 10\%$. Furthermore the  positions
and the normalizations of the Fe lines will be measured with an accuracy of few
\% and 15\%, respectively. 

We should be able to accumulate X-ray spectra of such a quality in the local
Universe (z$<$0.02) for Compton Thick AGN with $L_{10-40 keV} \sim 10^{43}$ \es
and up to z=0.2 for  Compton Thick  AGN with $L_{10-40 keV}$ $\sim 10^{45}$ \es
(see the line corresponding to $\sim 10^{-11}$ \cgs in Fig. \ref{fig1}).
According to the CXB synthesis models (see Comastri et al., these proceedings)
$\sim 40$ Compton Thick AGN at $|b|>20^o$ above  this flux limit are  expected.
Possible Simbol-X targets with this flux  are obviously the well known Compton
Thick AGN reported in  Table 1 as well as Infrared Bright Local Galaxies, that
can host Compton Thick AGN which are at the present optically elusive (so not
identified yet, see e.g. the case of Arp 299, \citealp{dellaceca02};
\citealp{ballo04}).

\subsection{$f_{10-40 keV} \sim 10^{-12}$ \cgs}

\begin{figure}[]
\resizebox{\hsize}{!}{\includegraphics[angle=-90,clip=true]{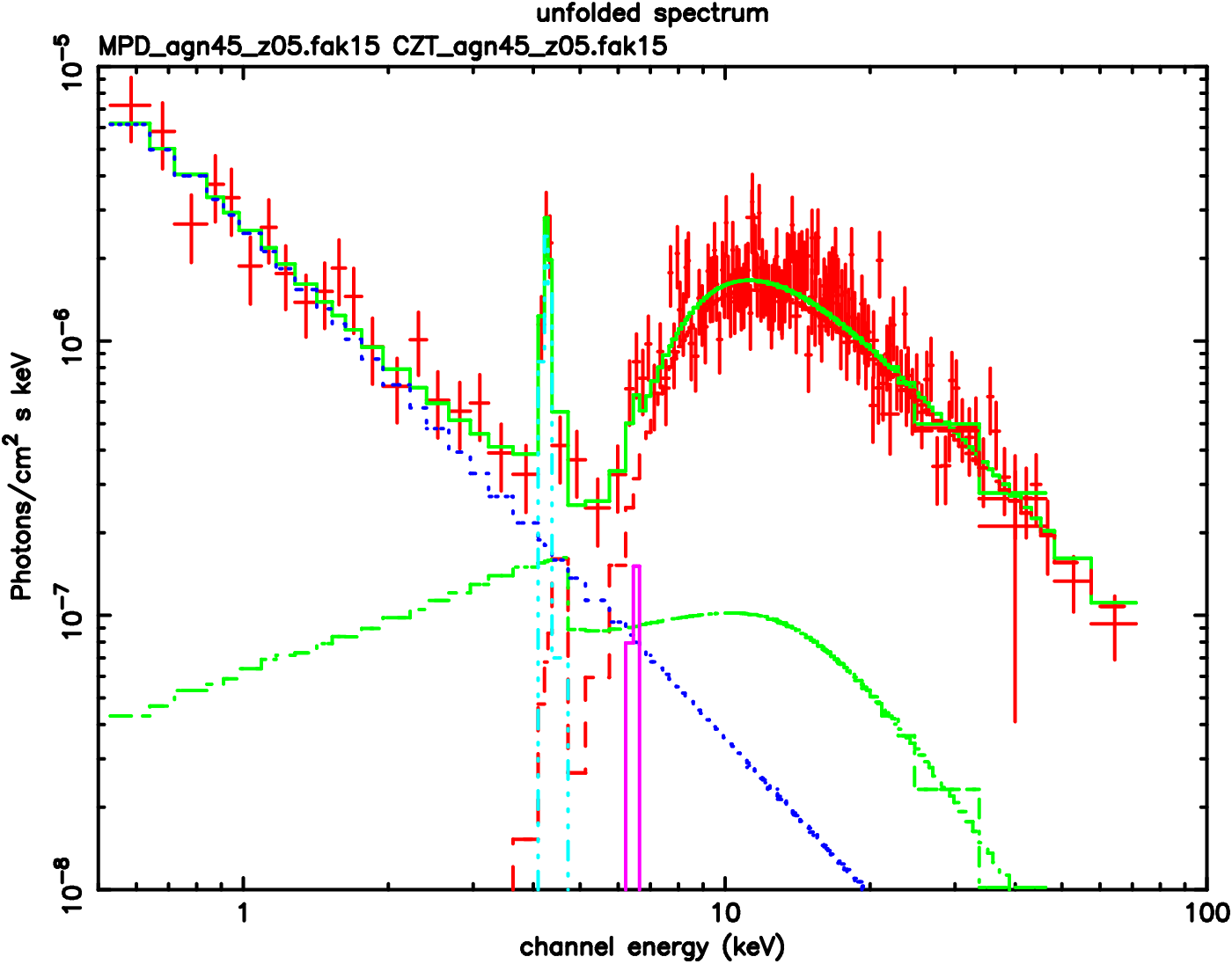}}
\caption{
\footnotesize
As figure 2 but for one object having $f_{10-40 keV} \sim 10^{-12}$ \cgs, 
corresponding to a source with a $L_{10-40 keV} \sim 10^{45}$ \es  at z$\sim$
0.5. 
}
\label{fig3}
\end{figure}

A 100 ksec simulation of a Compton Thick AGN with a $L_{10-40 keV} \sim 10^{45}$ \es at
z$\sim$ 0.5 is shown in Fig. \ref{fig3}  ($f_{10-40 keV} \sim 10^{-12}$ \cgs).
About $\sim$ 1500 MPD net counts and $\sim$ 800 CZT net counts will be
accumulated allowing us to determine the $N_H$ and the intrinsic 
luminosity with an accuracy of $\sim 10\%$ and $\sim 20\%$, respectively. The 
positions and the normalizations of the Fe lines will be measured with an
accuracy of few \% and $\sim 40\%$, respectively. 

X-ray spectra of this quality are expected for Compton Thick AGN with  $L_{10-40
keV}$ $\sim 10^{45}$ \es up to z$\sim$ 0.5, and  according to the CXB synthesis
modeling $\sim1500$ Compton Thick  AGN at $|b|>20^o$  are expected above  $\sim
10^{-12}$ \cgs. Possible Simbol-X targets are some of the candidate Compton
Thick  AGN listed in Table 2, Infrared Bright Local Galaxies and well known 
Ultra Luminous Infrared Galaxies (ULIRGS, see e.g. the case of IRAS 19254-7245,
\citealp{braito03}).

\subsection{$f_{10-40 keV} \sim 10^{-13}$ \cgs}

\begin{figure}[]
\resizebox{\hsize}{!}{\includegraphics[angle=-90,clip=true]{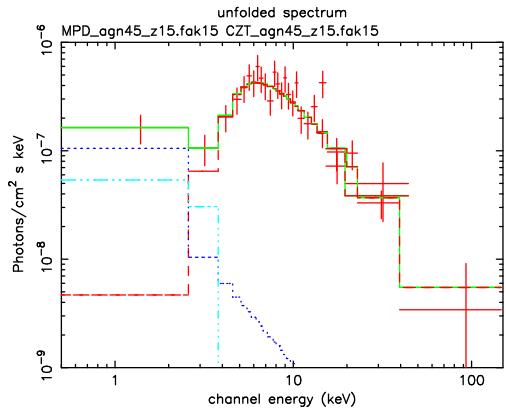}}
\caption{
\footnotesize
As figure 2 but for one object having $f_{10-40 keV} \sim 10^{-13}$ \cgs, 
corresponding to a source with a $L_{10-40 keV} \sim 10^{45}$ \es  at z$\sim$
1.5. 
}
\label{fig4}
\end{figure}

A Compton Thick AGN with a $L_{10-40 keV} \sim 10^{45}$ \es at z$\sim$ 1.5
should appear as shown in Fig. \ref{fig4}  ($f_{10-40 keV} \sim 10^{-13}$ \cgs;
$\sim$ 350 MPD net counts;  $\sim$ 150 CZT net counts). The intrinsic  $N_H$
will be determined with an accuracy $\sim 20-40\%$, while  the intrinsic
luminosity will be determined within a factor 2.

A few million of Compton Thick AGN ($|b|>20^o$)
are expected at this flux limit and possible 
Simbol-X targets are the candidate Compton Thick AGN selected from  the
Spitzer/Hershel  surveys (e.g.  \citealp{polletta06}, \citealp{daddi07},
\citealp{fiore07}) and some of the faintest ULIRGS. 

\subsection{Heavily Compton Thick AGN}

\begin{figure}[]
\resizebox{\hsize}{!}{\includegraphics[angle=-90,clip=true]{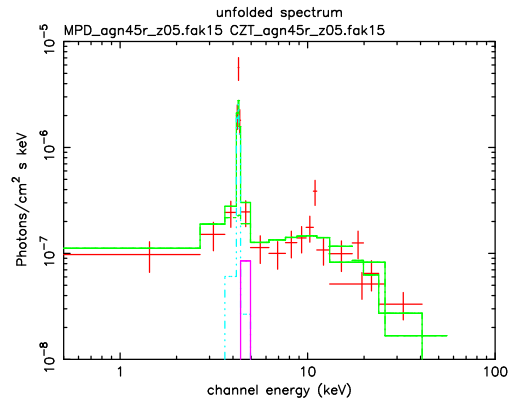}}
\caption{
\footnotesize
Simulated 100 ksec Simbol-X observation for a heavily Compton Thick AGN 
(spectral
model as described in the text) corresponding to a source with a  $L_{10-40 keV}
\sim 10^{45}$\es  at z$\sim$ 0.5.  
}
\label{fig5}
\end{figure}

Finally in Fig \ref{fig5} we show a simulated 100 ksec Simbol-X spectrum for a
heavily Compton Thick AGN   corresponding to a source with an intrinsic
luminosity   $L_{10-40 keV} \sim 10^{45}$\es at z$\sim$ 0.5  ($f_{10-40 keV}
\sim 7\times 10^{-14}$ \cgs; $\sim$ 250 MPD net counts;  $\sim$ 100 CZT net
counts).  Deeply  buried sources like NGC 1068 can be investigated up to
cosmological  distances. 

\section{Conclusions}

The Compton Thick AGN universe is practically a new field in high  energy
astrophysics since very few Compton Thick AGN (mostly in the local Universe)
have been  found and studied so far. The
Simbol-X mission, with its improved sensitivities  above 10 keV with respect to
previous satellites, is expected to open this  field to detailed investigation
up to cosmological distances ($z\sim 0.5-1.5$).

\begin{acknowledgements}
We thanks V. Braito and L. Ballo for useful discussions.
This work reveived partial financial support from ASI, MIUR (PRIN 
2006-02-5203) and INAF. 
\end{acknowledgements}

\bibliographystyle{aa}

\end{document}